# Hopping Conduction in Disordered Carbon Nanotubes

D. P. Wang*[1], D. E. Feldman,[1] B. R. Perkins,[2] A. J. Yin,[2] G. H. Wang,[1] J. M. Xu,[1,2] and A. Zaslavsky[1,2]

[1] Dept. of Physics, Brown University, Providence, RI 02912

[2] Div. of Engineering, Brown University, Providence, RI 02912

We report electrical transport measurements on individual disordered carbon nanotubes, grown catalytically in a nanoporous anodic aluminum oxide template. In both as-grown and annealed types of nanotubes, the low-field conductance shows as $\exp[-(T_0/T)^{1/2}]$ dependence on temperature $T$, suggesting that hopping conduction is the dominant transport mechanism, albeit with different disorder-related coefficients $T_0$. The field dependence of low-temperature conductance behaves an $\exp[-(\xi_0/\xi)^{1/2}]$ with high electric field $\xi$ at sufficiently low $T$. Finally, both annealed and unannealed nanotubes exhibit weak positive magnetoresistance at low $T = 1.7$ K. Comparison with theory indicates that our data are best explained by Coulomb-gap variable range hopping conduction and permits the extraction of disorder-dependent localization length and dielectric constant.

* Email: Dapeng_Wang@Brown.edu

I   INTRODUCTION

In recent years, the transport properties of one-dimensional (1D) or quasi-1D conductors moved into the focus of both fundamental and applied research. Quantum wires, nanofibers,[1,2] and especially carbon nanotubes (CNTs) have been the focus of intense scrutiny because of their possible incorporation in nanoelectronic devices. Extensive electrical transport studies have been carried out on various types of nanotubes, which can be metallic or semiconducting depending on chirality. The electronic transport properties of semiconducting CNTs vary from ballistic to diffusive, depending on the material quality.[3]

In an effort to realize large scale and self-assembled CNT arrays, nanotubes have been grown in nanoporous anodic aluminum oxide templates.[4,5,6,7] In this method, aluminum substrates are anodized under special conditions to produce regular pores with diameter of approximately 50 nm. After the deposition of a magnetic catalyst, these nanopore templates are exposed to carbon-carrying gases, such as acetylene, at 600 °C. The deposition of carbon material on the catalyst produces regular arrays of CNTs of nearly uniform ~50 nm outer diameter and ~35 nm inner diameter. These template-grown nanotubes are known to possess a significant amount of structural disorder,[7,8] in contrast with the nanotubes produced by laser ablation and arc discharge. The structural disorder of these nanotubes can change the dominant conduction mechanism from ballistic to hopping, in the geometry of a disordered 2D system wrapped into a cylinder. In the recent report by Jang and co-workers,[8] activated temperature dependence of conduction in disordered CNTs was analyzed in terms of 2D variable-range hopping.

In this paper, we report electrical transport measurements on individual template-grown CNTs, both as deposited and after a high temperature anneal, and interpret them in terms of Coulomb-gap variable range hopping (CGVRH). The conductance shows an $\exp[-(T_0/T)^{1/2}]$ temperature dependence at low electric field $\xi$ and $\exp[-(\xi_0/\xi)^{1/2}]$ dependence on the applied electric field $\xi$ at high field regime and low $T$. The hopping parameters $T_0$ and $\xi_0$ are determined by the disordered structure of the CNTs. Within the framework of hopping conduction theory, we relate $T_0$ and $\xi_0$ to the localization length. Finally, our choice CGVRH theory is corroborated by the weak low-temperature magnetoresistance of both annealed and unannealed CNTs. From the magnetoresistance, we extract the localization length, which agrees with the temperature and field dependence of the conductance, and the dielectric constant $\kappa$, which is consistent with previous studies of porous carbon structures.[9]

II DEVICE FABRICATON AND MEASUREMENTS

Our nanotubes were chemically removed from their growth template and dissolved in ethanol ultrasonically. Then the nanotubes were dispersed on a 500-nm-thick $SiO_2$-covered Si substrate prepatterned with alignment marks. Individual CNTs were located by scanning electron microscopy (SEM) and multiple contact patterns were patterned using e-beam lithography and Pd ohmic contact deposition. Subsequently, larger Ti/Au contact pads to the Pd contacts are patterned and deposited, allowing for a four-point measurement geometry shown in Fig. 1(a). Since previous experiments have shown that CNT electrical properties can improve upon a high-temperature anneal,[7, 10, 11] we annealed some of our CNTs at 1400 °C for 4 hours in flowing forming gas (Ar + 2% $H_2$).

Multilayer walls of the CNT and the disordered structure can be seen in Fig. 1(b), which shows a TEM image of an annealed CNT.

Four-point current-voltage $I(V)$ measurements were used to separate the contact resistance from the CNT itself. We find that for Pd contacts, the contact resistance is much smaller (<10 %) than the CNT resistance for all temperatures. By using the silicon substrate as the back gate, we can change the conductance of the nanotubes, proving the CNTs are $p$-type, as observed by other groups.[12] But the gating effect, which is very small for all the temperatures, demonstrates the deeply doped and highly defective nature of our CNTs.

The four-point current-voltage $I(V)$ characteristics of both annealed and as-deposited nanotubes at various temperatures $T = 1.7–300$ K are shown in Fig. 2. Figure 2(a) shows the $I(V)$ of an annealed CNT, which is 2.6 $\mu$m long between the central two electrodes. At room temperature, the resistance is linear and small. As $T$ is lowered, the resistance increases and for $T < 29$ K the resistance becomes nonlinear. At even lower temperature $T < 7.9$ K, the temperature dependence of the $I(V)$ characteristics saturates at high bias voltage, while the low bias resistance keeps. Figure 2(b) shows similar behavior for a 7 $\mu$m long unannealed CNT, except that the current is much smaller at the same voltage. When the temperature drops below 56 K, the resistance becomes nonlinear. At lower temperature $T < 11.8$ K, the low bias resistance keeps increasing until the current is comparable to the noise of the system (~10 pA).

We obtain the low bias conductance $\sigma$ by taking the derivative of $I(V)$ characteristics near $V = 0$. Figure 3 shows the low bias conductance as a function of temperature normalized by the conductance at $T = 300$ K ($\ln(\sigma(T)/\sigma(300$ K$))$ vs. $T^{-1/2}$) for both annealed and unannealed CNTs. For comparison, an alternative activated fit
(($\ln(\sigma(T)/\sigma(300 K))$) vs. $T^{-1}$), is shown as an inset in Fig. 3. Both the $T^{-1}$ fit and other
physically reasonable fits ( vs. $T^{-1/3}$ and vs. $T^{-1/4}$) are worse for both types of CNTs.
$\exp[-(T_0/T)^{1/2}]$ dependence has a much steeper slope (larger $T_0$) for unannealed CNTs.
We fabricated and measured four unannealed and two annealed samples. Although the
length of CNTs varied from 0.3 $\mu$m to 7 $\mu$m, the slope is consistent for each kind of CNT.
For the entire temperature range 1.7~300K, the change of the conductance of annealed
CNTs is less than two decades.

As a function of electric field $\xi$, the conductance of both types of nanotubes behaves
as $\exp[-(\xi_0/\xi)^{1/2}]$ at sufficiently high $\xi$ and low $T$. Figure 4 shows the electrical
conductance $\ln(\sigma/\sigma_0)$ plotted against $\xi^{-1/2}$ at various temperatures for both annealed
CNTs ($\sigma_0$ = 1 $\mu$S) and unannealed ($\sigma_0$ = 1 nS), where $\sigma_0$ is a normalization factor. For
annealed CNTs at $T$ = 1.8 K, the logarithm of conductance has a linear relation with
inverse square root of the electrical field, as $\xi$ > 0.006 V/$\mu$m (~ 0.015 V for a 2.6 $\mu$m long
device). The data are reasonably well described by an exponential dependence $\sigma(\xi)$ ~
$\exp[-(\xi_0/\xi)^{1/2}]$ where $\xi$ is the applied electric field. As the temperature increases, higher
electric field is needed to reach the high field regime, as seen in Fig. 4 (a). At $T$ = 4.2 K,
the high field regime is reached at $\xi$ > 0.02 V/$\mu$m, however at higher temperature, the
high field regime has not been reached with the bias voltage we applied. For unannealed
CNTs in Fig. 4(b), we see similar behavior, although the exponential dependence is
reached at higher $\xi$, and is characterized by a bigger $\xi_0$. At $T$ = 6.0 K, the high field
regime is reached at $\xi$ > 0.3 V/$\mu$m, which is 2 V for a 7 $\mu$m long device. The noise visible in Fig. 4(b) at the lowest temperature is due to the ~10 pA current noise floor of our measurement.

III MODELING

We now turn to interpreting our experimental observations of Figs. 2-4 in terms of Coulomb-gap variable range hopping conduction (CGVRH).[13] Previously, both power law and exponential $\exp[-(T_0/T)^{1/p}]$ temperature dependences of low-field conductance were observed in single and multiwalled CNTs. In case of SWNTs,[14,15] the behavior is consistent with the Luttinger liquid (LL) model which describes the correlated electronic state formed in a true one-dimensional system due to the Coulomb interaction. For MWNTs, the LL model has been investigated and can theoretically lead to a power-law behavior.[16] Experimentally, however, MWNTs are often shown to be diffusive conductors.[17] Langer and co-workers[18] have observed the conductance to exhibit a logarithmic temperature dependence and saturate at low temperature for an individual MWNT and used 2D weak localization (WL) theory to describe their data. For MWNTs grown in an AAO template, Davydov and co-workers[19] measured the tunneling spectra of a bundle of CNT arrays in the template and observed both $\exp[-T_0/T]$ and $\exp[-(T_0/T)^{1/2}]$ temperature dependence of the conductance, depending on the temperature range. A Y-junction CNT array grown in AAO was measured *in-situ* in the template by Papadopoulos and co-workers.[20] Power law dependencies were observed at a wide temperature range from 10~275K and interpreted partly in terms of the LL model. However, MWNTs are known to possess considerable disorder, making the LL model difficult to apply. Other recent experiments on either MWNT bundles or individual

MWNTs grown in AAO templates, reported $\exp[-(T_0/T)^{1/p}]$ temperature dependence of electrical conductance, with $p$ varying from 3 to 4.[8, 21]

In our experiment on individual MWNTs, we observe an exponential temperature dependence and field dependence of the conductance (Fig. 3 and 4), which can be explained by the CGVRH model, originally developed for the Miller-Abrahams resistor network.[22] In this model, the conductivity $\sigma$ between any pair of nodes, separated by $r_{ij}$ in the distance and $\varepsilon_{ij}$ in energy, behaves as

$$\sigma = \sigma_0 \exp\left(-\frac{\varepsilon_{ij}}{k_B T} - \frac{2r_{ij}}{\lambda}\right) \quad (1)$$

where $\sigma_0$ is some conductance prefactor and $\lambda$ is the localization length. The temperature dependence of the conductivity is given by

$$\sigma(T) = \sigma_0 \exp[-(T_0/T)^{1/2}] \quad (2)$$

with

$$T_0 = \frac{\beta e^2}{k_B \kappa \lambda} \quad (3)$$

where $\beta \cong 2.8$ is a dimensionless factor, derived by Shklovskii and Efros,[13] and $\kappa$ is the dielectric constant. A discussion of the CGVRH transport in granular carbon systems can be found in Ref. 9. It should be noted that the exponential temperature dependence of Eq. (2) is also compatible with one-dimensional VRH.[13] However, the fit based on 1D VRH leads to unreasonable predictions for the localization length.

Figure 3 plots $\ln(\sigma(T)/\sigma(300\ K))$ against $T^{-1/2}$ for both annealed and as-deposited CNTs, showing reasonable agreement with Eq. (2). Due to annealing, the material quality is highly improved. Thus, one also expects a longer localization length in the annealed case. As discussed below, this is compatible with the fact that the slope $T_0 = 36$ K for annealed CNTS is smaller than $T_0 = 900$ K for unannealed CNTs. These parameters will be used to deduce the localization length and dielectric constant later in this paper.

Transport becomes nonlinear at low temperatures and strong electric fields $\xi > k_B T/(e\lambda)$. We can understand nonlinear transport at zero temperature from the following argument in the frame of CGVRH theory. The energy cost of the electron transfer over the distance r exceeds $\varepsilon(r) \geq e^2/(\kappa r)$.[13] At zero temperature this energy can only be supplied by the external electric field. Thus, the electron transfer is only possible for $r > \varepsilon(r)/(e\xi)$, i.e. $r > \sqrt{e/(\kappa\xi)}$. The electric current is proportional to the probability of such transfers and hence to the square of the overlap matrix element between the localized states, with the latter scales as $\exp(-2r/\lambda)$. Hence, the low-temperature conductance can be expressed in the form

$$\sigma \propto \exp[-(\xi_0/\xi)^{1/2}] \qquad (5)$$

where $\xi_0$ is on the order of $4e/(\kappa\lambda^2)$.

Both types of CNTs we measured follow Eq. (5) at high $\xi$ and low $T$. Due to the relatively large localization length in annealed CNTs, they require smaller $\xi$ to meet the high electric field requirement at which nonlinear transport (5) occurs, as shown in Figs. 4(a) and (b). We find $\xi_0 \cong 0.1$ V/$\mu$m for annealed CNTs and 20.7 V/$\mu$m for as-deposited CNTs.

## IV MAGNETORESISTANCE

In order to corroborate our theoretical model, we measured the magnetoresistance of CNTs in transverse magnetic field at low $T$. Positive MR are observed, which can be explained by the field-induced shrinkage of the wavefunction. Accompanying the reduced wavefunction overlap is a decrease in the probability of tunneling. The temperature and the field dependence of the positive MR at small magnetic field are given by

$$R(B) = \exp[t\left(\frac{\lambda}{\eta}\right)^4 \left(\frac{T_0}{T}\right)^{3/2}] = \exp[\alpha B^2] \qquad (6)$$

where $\eta = \sqrt{\hbar/eB}$ is the magnetic length and $t \sim 0.0015$ is a numerical factor, derived in Ref 13. Figure 5(a) shows the normalized magnetoresistance $(R(B)-R_0)/R_0$ as a function of the magnetic field $B$ for annealed CNTs, where $R_0$ is the resistance at $B = 0$. CGVRH theory predicts $\ln(R) \sim B^2$ at low fields such that magnetic length $\eta = \sqrt{\hbar/eB} \geq \lambda$. In our case, the critical field is $B_c^2 \sim (\frac{\hbar}{e\lambda^2})^2 \sim 10$ T$^2$ for $\lambda = 15$ nm. Our data at both temperatures, plotted as $\ln(R(B)/R_0)$ vs. $B^2$ in Fig. 5(b), show a clear $\ln(R) = \alpha B^2$ dependence at low $B$, below the characteristic field $B_c$, which agrees with the CGVRH theory, with $\alpha = 5.6 \times 10^{-3}$ (T$^{-2}$) at 3.6 K and $\alpha = 2 \times 10^{-2}$ (T$^{-2}$) at 1.7 K for annealed CNTs. The localization length extracted from $\alpha$ at both temperatures is consistent. At strong fields, the theory[13] predicts a slow field dependence of the magnetoresistance, $\ln(R) \sim B^{1/5}$ which qualitatively agrees with the saturation observed in the $T = 1.7$ K curve. The strong field saturation can emerge when the hopping distance is much greater than $\eta^2/\lambda$. Since the hopping distance is greater at lower temperatures,[13] lower temperatures correspond to lower saturation fields, which agrees with the fact that we did not see saturation in the $T = 3.6$ K curve. We also observed similar MR curves with the quadratic field dependence but a lower magnitude for unannealed CNTs, with $\alpha = 4.9\times10^{-3}$ (T$^{-2}$) at 4.0 K and $\alpha = 1.8\times10^{-2}$ (T$^{-2}$) at 1.7 K. The change of $\alpha$ with the temperature agrees with Eq. (6). From these values of $\alpha$ and $T_0$, we can deduce the localization length, which is $\lambda_{an}$ = 15 nm for annealed CNTs and $\lambda_{un}$ = 5 nm for as-deposited CNTs. Using Eq. (3), we find that $\kappa_{un} \cong$ 11, which is in the range of the dielectric constant of graphite, whereas $\kappa_{an} \cong 80$, which is similar to the results obtained for heat-treated porous carbon materials by Fung and co-workers.[9] Following Eq. (5), we can estimate $\xi_0$ = 0.3 V/$\mu$m for annealed CNTs and 20 V/$\mu$m for unannealed CNTs, which is in reasonable agreement with the experiment. A summary of the parameters we obtain from our analysis is presented in Table I.

## V CONCLUSION

In summary, we report electrical transport measurements of individual carbon nanotubes grown catalytically in a nanoporous anodic aluminum oxide template. The low-field conductance shows an $\exp[-(T_0/T)^{1/2}]$ temperature dependence, whereas the high-field conductance exhibits an $\exp[-(\xi_0/\xi)^{1/2}]$ dependence on electric field $\xi$ at low temperatures in both as-deposited and annealed CNTs. This suggests that Coulomb-gap variable range hopping is the dominant transport mechanism at low $T$. The change in localization length of the CNT material upon annealing agrees with the observed changes in the hopping parameter $T_0$. The relatively weak magnetoresistance in both as-deposited and annealed CNTs, following the $\exp[\alpha B^2]$ dependence on magnetic field $B$ at low fields, also agrees with the predictions of the CGVRH theory.

This work was supported by NSF CCF-0403958, NSF DMR-0544116, NSF ECS-0223943, ONR and the Salomon Award program at Brown University. The authors acknowledge the use of the Microelectronics Central Facility at Brown, supported by the NSF MRSEC (DMR-0079964).

TABLE CAPTIONS

Table I. Various physical parameters for as-deposited and annealed CNTs, including the experimental value of $T_0$, $\xi_0$, and $\alpha$, together with localization length and dielectric constant calculated by the CGVRH theory.

|  | As-deposited CNTs | Annealed CNTs |
|---|---|---|
| $T_0$ (K) | 900 | 36 |
| $\alpha$ (T$^{-2}$) at 1.7 K | $1.8 \times 10^{-2}$ | $2 \times 10^{-2}$ |
| $\xi_0$ at the lowest $T$ (V/$\mu$m) | 34 | 0.1 |
| $\lambda$ (nm) | 4.5 | 15 |
| $\kappa$ | 10.8 | 80 |

Table I: D. P. Wang *et al.*

FIGURE CAPTIONS

Fig. 1: (a) Low magnification scanning electron microscopy (SEM) image of an individual CNT with four Pd electrodes, showing the four-point measurement geometry (the scale bar is 1 $\mu$m); (b) TEM image of an annealed CNT ( the scale bar is 5 nm).

Fig. 2: Current-voltage $I(V)$ characteristics of individual template-grown CNTs measured at various temperatures $T$: (a) annealed CNTs, (b) unannealed CNTs.

Fig. 3: Temperature dependence of the conductivity of template-grown CNT material at zero bias, plotted as $\ln(\sigma(T)/\sigma(300\,K))$ vs. $T^{-1/2}$ for both annealed and unannealed CNTs. Inset shows alternative activation fitting of $\ln(\sigma(T)/\sigma(300\,K))$ vs. $T^{-1}$.

Fig. 4: The high field electrical conductivity $\ln(\sigma/\sigma_0)$ measured at various temperatures as a function of square root of the reciprocal electric field $\xi^{-1/2}$ for annealed CNTs (a) and unannealed CNTs (b) at various $T$ ($\sigma_0$ is a prefactor described in the text).

Fig. 5: (a) Normalized magnetoresistance $(R(B) - R_0)/R_0$ for annealed CNTs at different temperatures, where $R_0$ is the $B = 0$ resistance; (b) the same data plotted as $\ln(R(B)/R_0)$ vs. $B^2$, with the linear fit of low field regime for $T = 1.7$ K.

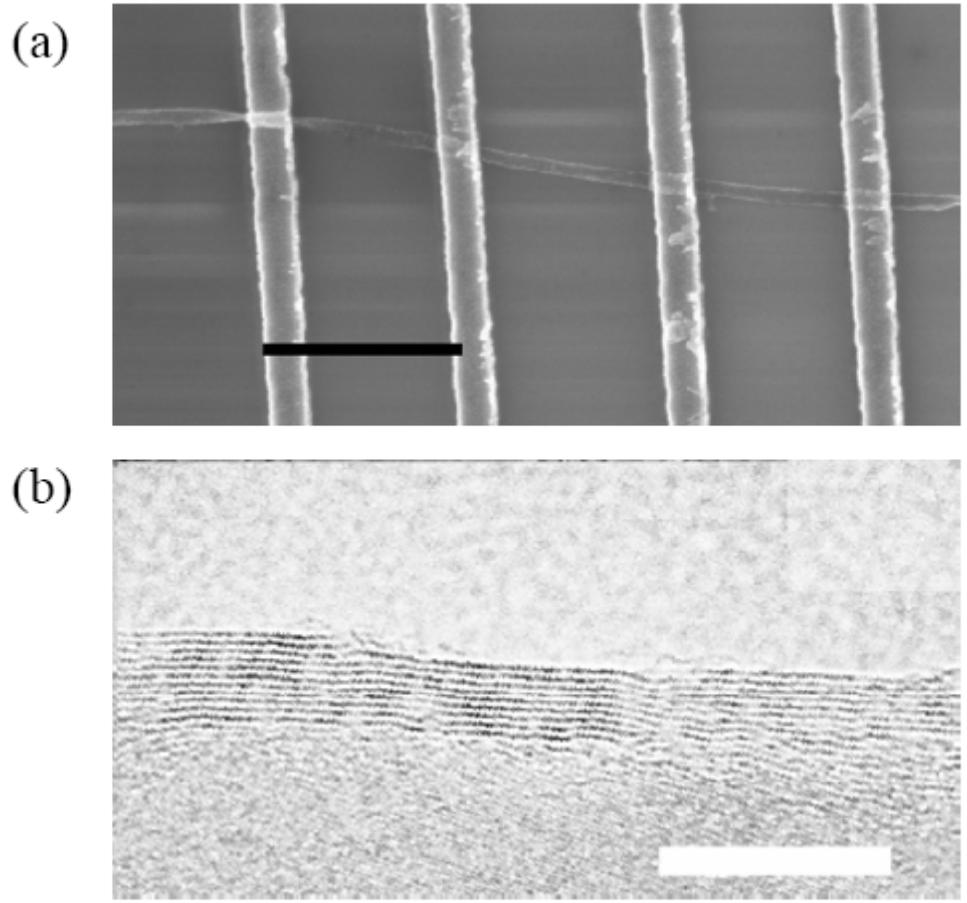

Fig. 1: D. P. Wang *et al.*

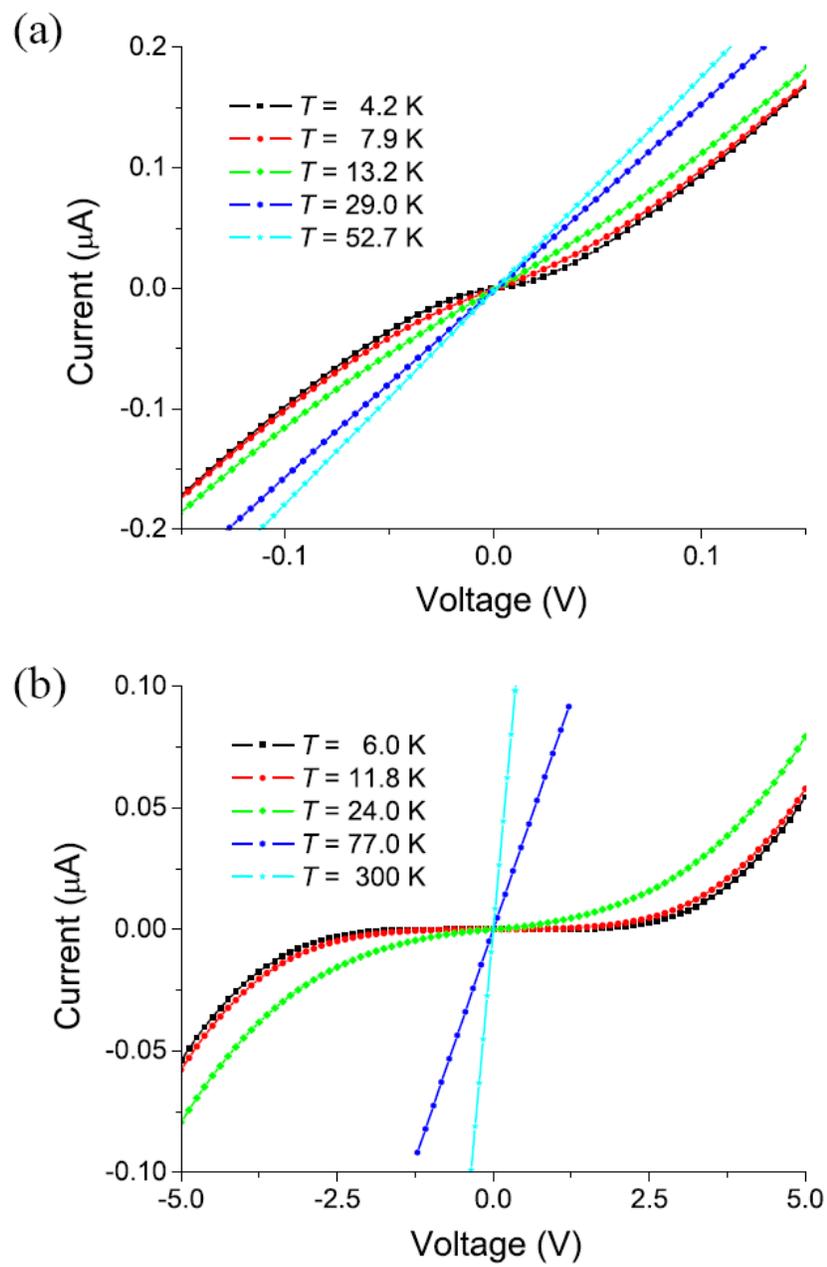

Fig. 2: D. P. Wang *et al.*

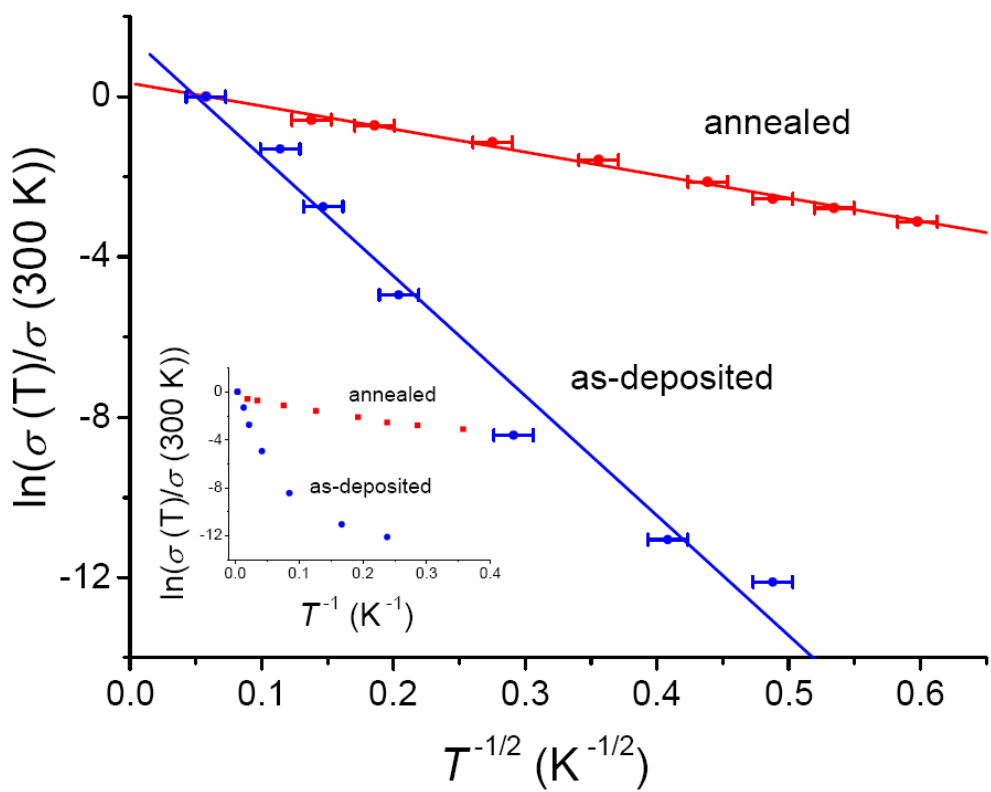

Fig. 3: D. P. Wang *et al.*

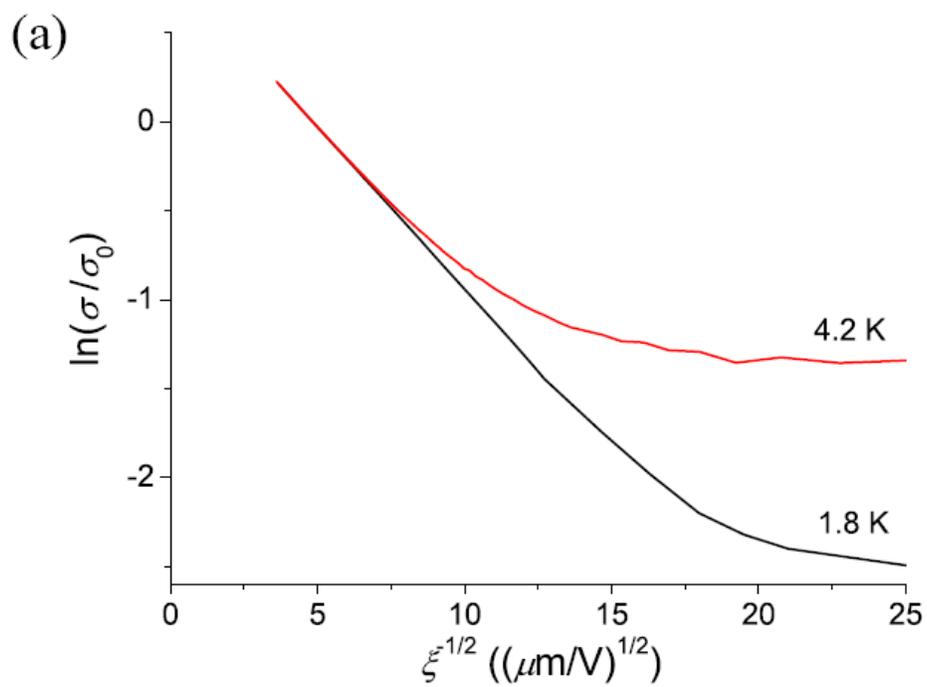
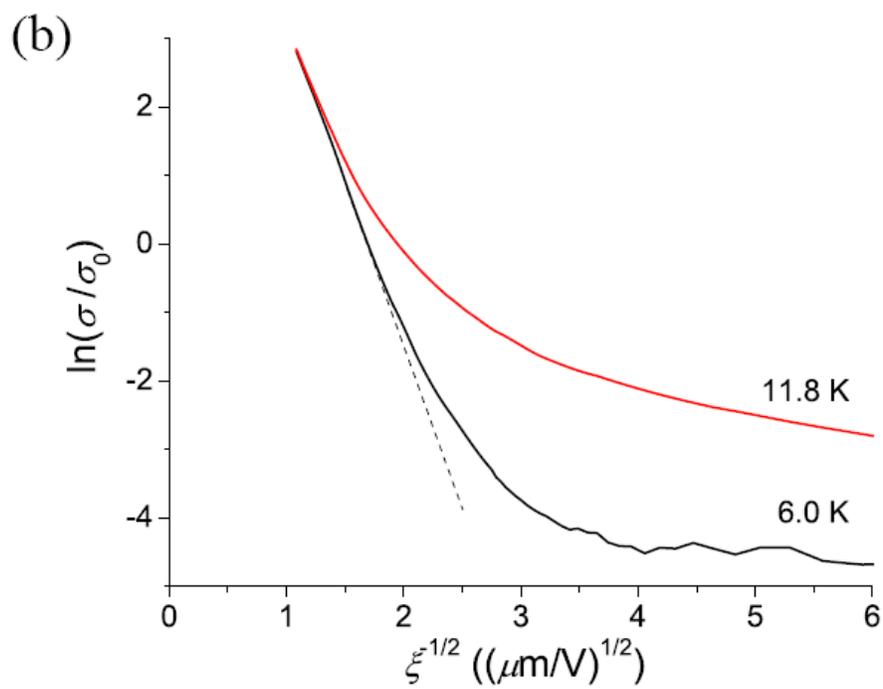

Fig. 4: D. P. Wang *et al.*

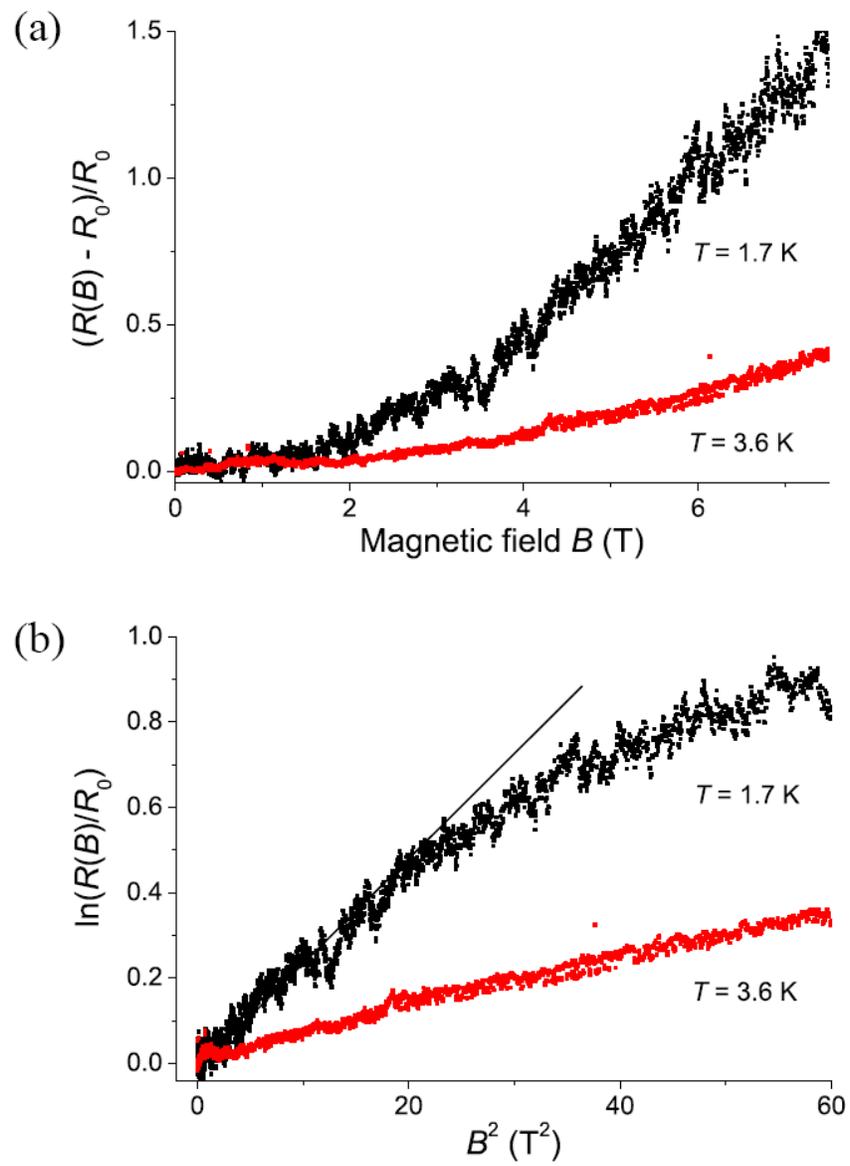

Fig. 5: D. P. Wang *et al.*